\documentstyle[aps,preprint,tighten,psfig]{revtex}

\def\bea{\begin{eqnarray}}
\def\eea{\end{eqnarray}}
\def\fr{\frac}
\def\l{\left}

\def\eps{\epsilon}

\def\eeas{\end{eqnarray*}}
\def\beas{\begin{eqnarray*}}
\def\ee{\end{equation}}
\def\be{\begin{equation}}

\def\veps{\varepsilon}

\def\fpi2{\mbox{F$_\pi$}^2}

\def\mpi2{{m_\pi}^2}
\def\mk{m_K}
\def\mk2{{m_K}^2}

\def\fk2{\mbox{F$_K$}^2}

\begin{document}

\draft

\preprint{TAN-FNT-00-03}

\title{Multibaryons in the collective coordinate approach to the SU(3) Skyrme model}

\author{C. L. Schat \thanks{Email address: schat@cbpf.br} $^{\rm 1}$ and 
N. N. Scoccola \thanks{Email address: scoccola@tandar.cnea.gov.ar} $^{\rm
2,3,4}$}

\address{$^{\rm 1}$ Instituto de F\'{\i}sica - DFNAE,
Universidade do Estado do Rio de Janeiro, \\
Rua S\~ao Francisco Xavier 524, Maracan\~a, 20559-900 Rio de Janeiro, RJ,
Brazil.}

\address{$^{\rm 2}$ Consejo Nacional de Investigaciones 
Cient{\'\i}ficas y T\'ecnicas, Argentina.}

\address{$^{\rm 3}$ Departamento de F{\'{\i}}sica, 
CNEA, Avda. Libertador 8250, (1429) Buenos Aires, Argentina.}

\address{$^{\rm 4}$ Universidad Favaloro, Sol{\'\i}s 453, 
(1078) Buenos Aires, Argentina.}

\maketitle
\vspace*{1cm}
\centerline{\today}
\begin{abstract}
\noindent We obtain the rotational spectrum of strange multibaryon states by performing
the SU(3) collective coordinate quantization of the static multi-Skyrmions. These background
configurations are given in terms of rational maps, which  are very good approximations and
share the same symmetries as the exact solutions. Thus,  the allowed quantum numbers
in the spectra and the structure of the collective Hamiltonians we obtain are also
valid in the exact case. We find that the predicted spectra are in overall agreement with those
corresponding to the alternative bound state soliton model.
\end{abstract}



\section{Introduction}

In the Skyrme model\cite{Sky61} and its generalizations, baryons arise as topological
excitations of a non-linear chiral Lagrangian written in terms of meson fields. These  type of models
have been quite successful in describing the properties of single baryons such as the nucleon
and
the strange hyperons (see e.g., Refs.\cite{ZB86,Wei96}). This has lead people to investigate
the lowest energy Skyrmion configurations with topological number greater than one, which are of inherent
interest as examples of three dimensional solitonic structures and may also be
relevant for nuclear physics. These studies were
already started by Skyrme in his pioneer papers at the beginning of the sixties. However, it was
only in 1987 that the minimum energy $B=2$ Skyrmion was correctly identified\cite{KS87}. Some time
later the authors of Ref.\cite{BTC90} found the solutions with $B=3,4$ and $5$ by numerical relaxation calculations. Finally, a few years
ago\cite{BS97}, after some demanding numerical work,  the global minimum energy configurations with topological number up to $B=9$ were constructed. One
particularly interesting aspect of all these multi-Skyrmion fields is that they are very symmetric. While
for $B=2$ the symmetry group corresponds to that of a torus, for $B=3,4,7,9$ they possess
the symmetries of the platonic polihedra $T_d, O_h, I_h$ and $T_d$, respectively, and for
$B=5,6,8$ the dihedral symmetries $D_{2d},D_{4d},D_{6d}$, respectively.  It should be stressed that, in spite of this,
the multi-Skyrmion fields are very complicated functions of the space coordinates which are only
known numerically. Fortunately, rather simple and accurate approximations to these configurations have
been found\cite{HMS98}. They are based on some {\it Ans\"atze} which are written in terms of rational maps
and take advantage of the similarities between multi-Skyrmion fields and
Bogomol'nyi-Prasad-Sommerfield (BPS) monopoles.
These developments triggered several investigations concerning the properties of the
multi-Skyrmions (such as e.g., vibrational excitations\cite{BBT97}) as well as their application to
baryonic systems containing strangeness\cite{SS98,GSS00} and heavier flavors\cite{SS00}. The extension to flavored multibaryons is also
motivated by the advent of heavy ion colliders with the possibility of producing strange\cite{E864} and even charmed\cite{SV98} multibaryonic states with rather low baryon
number in the laboratory.
To describe the strange multibaryons one has to extend the model to
SU(3) flavor space. The classical background configurations are simply
obtained by embedding the SU(2) static multi-Skyrmions in the
isospin subgroup of SU(3). In order to obtain the spectrum
with states of well defined spin and isospin quantum numbers,
as well as their splittings, we have to perform the quantization of
this system. However,  the presence of the rather important
symmetry breaking terms associated with the mass of the strange quark
makes the quantization process not completely trivial. In fact, two alternative
methods have been suggested in the literature. One is known as the
 bound state approach\cite{CK85} (BSA) in which strange baryons
are described as SU(2) rotating soliton-kaon bound systems.
The other scheme assumes that the strange degrees of freedom can
still be treated as rotational modes but the corresponding
collective Hamiltonian is to be diagonalized exactly\cite{YA88}. This method
is usually called the rigid rotator approach (RRA). In
two recent articles\cite{SS98,GSS00} SU(3) multi-Skyrmions have
been investigated following the BSA. In this work we complement
such investigations by considering these configurations within the
framework of the RRA.

This paper is organized as follows. In Sec. II we provide a brief description
of the model and obtain the collective Hamiltonian for the different baryon numbers.
In Sec. III we focus on the determination of the multibaryon quantum numbers and
wavefunctions. In Sec. IV we present the numerical results and in Sec. V our conclusions.
Finally, in the Appendix we give the explicit form of the collective Hamiltonians for
$3 \le B \le 9$.

\section{The Model}

We start with the effective action of the SU(3) Skyrme model supplemented with an
appropriate symmetry breaking term \cite{Wei96}. Expressed in terms of the
SU(3)-valued chiral field $U(x)$ it reads
\bea
\Gamma = \int d^4x \ \left\{
{f^2_\pi \over{4}} {\rm Tr}\left[ \partial_\mu U \partial^\mu U^\dagger \right] +
{1\over{32 e^2}} {\rm Tr}\left[ [U^\dagger \partial_\mu U ,
U^\dagger \partial_\nu U ]^2 \right] \right\} + \Gamma_{WZ} +
\Gamma_{SB} \ ,  \label{action}
\eea
where $f_\pi$ is the pion decay constant (~$= 93 \ \rm{MeV}$ empirically)
and $e$ is the so-called Skyrme parameter.  In Eq. (\ref{action}),
the symmetry breaking term $\Gamma_{SB}$ accounts for the different
masses and decay constants of the pion and kaon fields while
$\Gamma_{WZ}$ is the usual Wess-Zumino action. Their explicit
forms are
\bea
\Gamma_{SB}
& \! \! = \! \! &\int d^4x \left\{ {f_\pi^2  m_\pi^2 + 2 f_K^2 m_K^2 \over{12}}
{\rm Tr} \left[ U + U^\dagger - 2 \right]
+ { f_\pi^2 m_\pi^2 - f_K^2 m_K^2 \over{6}}
{\rm Tr} \left[ \sqrt{3} \lambda^8 \left( U + U^\dagger \right) \right] \right.
\nonumber \\
& & \hskip 1.cm \left. + { f_K^2 - f_\pi^2 \over{12} } {\rm Tr}
\left[ \left( 1 - \sqrt{3} \lambda^8 \right)
\left( U \partial_\mu U^\dagger \partial^\mu U +
        U^\dagger \partial_\mu U \partial^\mu U^\dagger \right) \right]
\right\} \ ,
  \\
\Gamma_{WZ} \ &=& \ -i \fr{N_c}{240\pi^2}\int \ d^5x \
\veps^{\mu\nu\alpha\beta\gamma}
\ {\rm Tr}\l[ U^\dagger (\partial_\mu U) U^\dagger (\partial_\nu U) U^\dagger (\partial_\alpha U) U^\dagger (\partial_\beta U) U^\dagger (\partial_\gamma U) \right] \ ,
\label{oldmass}
\eea
where $\lambda^8$ is the eighth Gell-Mann matrix, $N_c$ the number of colors,
 $m_\pi$ and $m_K$
are the pion and kaon masses, respectively,
and $f_K$ is the kaon decay
constant.

We proceed by introducing the following {\it Ansatz} for the
time dependent chiral field
\begin{equation}
U({\mbox{\boldmath $\vec r$}}, t) = {\cal A}(t) \ \left(
\begin{array}{cc}
\exp \left[ i \mbox{\boldmath $\vec \tau$} \cdot
{{\mbox{\boldmath $\vec \pi$}}}({\cal R}^{-1}(t) {{\mbox{\boldmath $\vec r$}}}) \right] & 0 \\ 0 & 1
\end{array}
\right) \ {\cal A}^\dagger(t) \ ,
\label{ansatz}
\end{equation}
where the embedded SU(2) background configuration is rigidly rotated
both in SU(3) flavor space and real space, the collective coordinates
given by ${\cal A}(t)\in$ SU(3) and ${\cal R}(t)\in$ SO(3), respectively. Substituting $U(\mbox{\boldmath $\vec r$}, t) $ given by Eq.\ (\ref{ansatz})
into the effective action yields a Lagrangian of the general form
\begin{equation}
L = - M_{sol} + L_{coll} \ ,
\end{equation}
where $M_{sol}$ is the static SU(2) soliton mass and $L_{coll}$ is the collective
Lagrangian, whose general expression will be given below. Following the usual steps in the
RRA,  we first find the soliton background
configuration by minimizing $M_{sol}$. For this purpose we introduce the rational map {\it Ans\"atze} \cite{HMS98} for the pion field
\be
{{\mbox{\boldmath $\vec \pi$}}} ({{\mbox{\boldmath $ \vec r$}}}) =
 F(r)  \ {\hat{\mbox{\boldmath $n$}}} \ .
\label{mansatz}
\ee
Here, $F(r)$ is the multi-Skyrmion profile which depends on the radial coordinate
only and $\mbox{\boldmath $\hat n$}$ is a unit vector given by
\be
{\hat{\mbox{\boldmath $n$}}} = {1\over{1+|R|^2}}
\left[ 2 \ \Re(R) \ {\hat{\mbox{\boldmath $\imath$}}} +
       2 \ \Im(R) \ {\hat{\mbox{\boldmath $\jmath$}}} +
    ( 1 - |R|^2 ) \ {\hat{\mbox{\boldmath $k$}}} \right] \ ,
\label{pians}
\ee
with $R = R(z)$
the rational map corresponding to a certain winding number $B$ which is identified with the baryon number.
The
complex variable $z$ is related to the usual two spherical
coordinates $(\theta,\phi)$ via stereographic
projection, namely, $z = \tan(\theta/2) \exp(i \phi)$.
For example, the map corresponding to the $B=1$ hedgehog {\it Ansatz}
is the identity map $R = z$. The explicit form of the rational maps corresponding
to the other baryon numbers $B \leq 9$
and the resulting expression for the soliton mass $M_{sol}$
can be found in Refs.\cite{HMS98,GSS00}. The radial profile function $F(r)$ is determined by minimizing the classical soliton
energy $M_{sol}$. Details of this procedure as well as plots of these
profiles for different baryon numbers are given in Ref.\cite{HMS98}.

The collective Lagrangian
written in terms of the collective degrees
of freedom and the corresponding angular velocities $\Omega_a, \omega_\alpha$ defined by
\footnote{Here and in the following the spin/isospin indices $a,b,c$ run over $\{1,...,3\}$, the flavor index $\alpha $ over  $\{1,...,8\}$ and the  $k \in \{4,...,7\}$ index  corresponds to excitations into strangeness directions. }
\bea
\label{angvel}
\left ( {\cal R}^{-1} \dot {\cal R} \right )_{a b} &=& \eps_{a b c}\Omega_{c} \ , \\
{\cal A}^{-1} \dot {\cal A} &=& \fr{i}{2}\  \lambda_\alpha \ \omega_\alpha \ ,
\eea
takes the general form
\begin{eqnarray}
L_{coll} & = & {1\over2}  \sum_{a,b} \left( \Theta^J_{a b} \ \Omega_a \Omega_{b} +
    \Theta^N_{a b} \ \omega_a \omega_{b} +
    2 \ \Theta^M_{ab} \ \Omega_a \omega_{b} \right) +
\frac{1}{2} \Theta^S \sum_{k} \omega_k^2  - {N_c B\over{2 \sqrt{3}}} \ \omega_8  \nonumber \\
& & \qquad \qquad - {1\over2} G_{SB} \left( 1 - D_{88} \right) \ ,
\label{lag}
\end{eqnarray}
with $D_{88} = \fr{1}{2} \ {\rm Tr} \left[ \lambda_8 {\cal A} \lambda_8 {\cal A}^\dagger \right]$. The moment of inertia in the strangeness direction $\Theta^S$ is
\begin{equation}
\Theta^S = \int d^3r \ { 1 - c\over2}
\left[ f_K^2 + \frac{1}{4 e^2 } \left( {F'}^2 + 2 B \fr{s^2}{r^2}\right) \right] \ ,
\end{equation}
where we have introduced the short hand notation $s = \sin F$, $c = \cos F$.
The spin $\Theta^J_{ab}$, isospin $\Theta^N_{ab}$ and mixed moments of inertia $\Theta^M_{ab}$ are
\bea
\Theta^J_{a b} &=& \ \ \int d^3r \ s^2 \ r^2 \ \left[ \left( f_\pi^2  + {F'^2 \over{e^2}} \right) +
\frac{1}{2} \ {s^2\over{e^2}} \
\nabla_{c} {\hat{\mbox{\boldmath $n$}}}\cdot \nabla_{c} {\hat{\mbox{\boldmath $n$}}}
 \right] \ \nabla_a {\hat{\mbox{\boldmath $n$}}}\cdot \nabla_{b}
{\hat{\mbox{\boldmath $n$}}} \
 ,
\label{tjab}
\\
\Theta^N_{a b}&=& \ \int d^3r \  s^2 \ \left[\left( f_\pi^2  + {F'^2 \over{e^2}} \right)(\delta_{a b}- \hat{\mbox{$n$}}_a \hat{\mbox{$n$}}_{b}) \
 + {s^2\over{e^2}}
 \ (\delta_{a b}- 2 \hat{\mbox{$n$}}_a \hat{\mbox{$n$}}_{b})   \
\nabla_{c} {\hat{\mbox{\boldmath $n$}}}\cdot \nabla_{c} {\hat{\mbox{\boldmath $n$}}} \right] \ ,
\label{tiab} \\
\Theta^M_{a b} &=& - \int d^3r \ s^2 \ r \ \left[ \left( f_\pi^2  + {F'^2 \over{e^2}} \right) +
\frac{1}{2} \ {s^2\over{e^2}}  \
\nabla_{c} {\hat{\mbox{\boldmath $n$}}}\cdot \nabla_{c} {\hat{\mbox{\boldmath $n$}}}
\right] \  \nabla_a \hat{\mbox{$n$}}_{b} \
 \  .
\eea
For the rational maps we are interested in
 all these moments of inertia are diagonal \cite{GSS00}.
This is a direct consequence of the symmetries of these {\it Ans\"atze}.
 Finally, the symmetry breaking parameter $G_{SB}$ is
\begin{equation}
G_{SB} = \frac{2}{3} (f_K^2 - f_\pi^2) \int d^3r  \left(   {F'}^2 +  2 B \fr{s^2}{r^2} \right) c +
 \frac{4}{3} (f_K^2 m_K^2 - f_\pi^2 m_\pi^2 ) \int d^3r \ (1-c) \ .
\end{equation}

Given $L_{coll}$, the spin and flavor canonical momentum operators $\hat J_a$ and $\hat F_\alpha$ are defined in the usual way
\begin{eqnarray}
\label{momenta}
\hat J_a = \frac{\partial L_{coll}}{\partial \Omega_a} \qquad ; \qquad
\hat F_\alpha = \frac{\partial L_{coll}}{\partial \omega_\alpha} \ .
\end{eqnarray}
The collective Hamiltonian is conventionally obtained as the Legendre transformation
$ H_{coll} = J_a \Omega_a + F_\alpha \omega_\alpha - L_{coll} $, resulting in
\begin{equation}
H_{coll} = K^S \left[ C_2(SU(3)) - {3\over4} B^2 - \hat N^2 + \gamma (1-D_{88}) \right] + H^{JN}_{B} \
. \label{nonvan}
\end{equation}
Here, $C_2(SU(3))= \sum_\alpha \hat F_\alpha^2$ stands for the quadratic SU(3) Casimir operator, $\hat N_a \equiv \hat F_a$ is the isospin operator in the soliton frame,
$\gamma = \Theta^S G_{SB}$ is the
dimensionless flavor symmetry breaking parameter and $K^S = 1/(2 \Theta^S)$.
In order to obtain  Eq. (\ref{nonvan}) we have used $N_c=3$ and
the constraint $F_8 = - {\sqrt{3}\over2} B$.
Finally,
the detailed form of the spin-isospin collective Hamiltonians $H^{JN}_{B}$ depends on the soliton symmetry group. The method to derive them
is very similar to the one described in Sec. III of Ref.\cite{GSS00}.
For $B=1, 2$ the corresponding groups are the continuous groups $O(3)$ and $D_{\infty h}$, respectively.
In those cases there are some relations between the spin and isospin operators which lead
to the  well known expressions for the spin-isospin collective Hamiltonians
\begin{eqnarray}
H^{JN}_{B=1} &=& \frac{1}{2 \Theta^J}  \hat J^2 \ ,
    \label{hb1} \\
H^{JN}_{B=2} &=&
    {1\over{2 \Theta_1^J}} \left(\hat J^2 - \hat J_3^2\right) +
    {1\over{2 \Theta_1^N}} \left(\hat N^2 - \hat N_3^2\right)
    + \frac{1}{2 \Theta_3^N} \ \hat N_3^2 \ .
    \label{hb2}
\end{eqnarray}
On the other hand, for $B \ge 3$ the symmetry groups are finite\cite{BS97}.
Thus, the general form of the spin-isospin collective Hamiltonian
is
\begin{eqnarray}
H^{JN}_{B \ge 3} &=& \sum_{a} \left( K_a^J  \ \hat J_a^2 + K_a^N \ \hat N_a^2 -
         2 K_a^M \ \hat J_a \ \hat N_a \right) \ ,
    \label{ha}
\end{eqnarray}
where
\begin{equation}
K_a^J =  \frac{1}{2} \frac{\Theta_a^N}{\Delta_a} \ ,  \quad
K_a^N =  \frac{1}{2} \frac{\Theta_a^J}{\Delta_a} \ ,  \quad
K_a^M =  \frac{1}{2} \frac{\Theta_a^M}{\Delta_a} \ .
\end{equation}
and $\Delta_a \equiv \Theta_a^J \Theta_a^N - (\Theta^M_a)^2$. The explicit form for each baryon number
can be found in the Appendix.

\section{Quantum numbers and collective wave functions}

In order to calculate the rotational corrections to the multi-Skyrmion masses we have to find the
corresponding wave functions. Following the Yabu-Ando procedure \cite{YA88},  they should
diagonalize the flavor symmetry breaking term in the collective Hamiltonian. At the same time
they should satisfy the constraints imposed by the symmetries of the classical soliton configuration.
Thus, the general form of such eigenfunctions will be
\begin{equation}
|B J J_z, Y I I_z, N \rangle = \sum_{J_3 N_3} \,
    \alpha^{JN}_{J_3 N_3} \, D^J_{J_z J_3} \, \Psi_{(Y, I, I_z),(B, N, N_3)} \ ,
    \label{jii}
\end{equation}
Here, $D^J_{J_z J_3}$ is the usual SU(2) Wigner function and $\Psi_{(Y, I, I_z),(B, N, N_3)}$ is
a function depending on the 8 Euler angles that parametrize the SU(3) manifold.
To obtain $\Psi_{(Y, I, I_z),(B, N, N_3)}$ we should solve the eigenvalue equation
\begin{equation}
K^S \left[ h + \gamma (1-D_{88}) \right] \ \Psi
= \epsilon \ \Psi \ ,
\label{eigen}
\end{equation}
where $h = C_2(SU(3)) - {3\over4} B^2 - N (N + 1)$. The coefficients $\alpha^{JN}_{J_3 N_3}$ are
determined in such a way that the full wavefunction transforms as  some particular one-dimensional
irreducible representation (irrep) of the soliton symmetry group $G$. This will be discussed in
some detail below.

To solve Eq. (\ref{eigen}) we expand $\Psi_{(Y, I, I_z),(B, N, N_3)}$ in a basis of SU(3) Wigner functions
$D^{(p,q)}_{(Y, I, I_z),(B, N, N_3)}$, where $(p,q)$ are the labels used to identify the SU(3) irrep.
Namely,
\begin{equation}
\Psi_{(Y, I, I_z),(B, N, N_3)} = \sum_{(p,q)} \beta_{(p,q)} \ \sqrt{d_{(p,q)}}
 \ D^{(p,q)}_{(Y, I, I_z),(B, N, N_3)} \ ,
\label{expansion}
\end{equation}
where $d_{(p,q)} = (p + 1) (q + 1) (p+q+2) /2$ is the dimension of the irrep.
In such basis $h$ is diagonal and the matrix elements of the symmetry breaking term can be
expressed as a product of two SU(3) Clebsch-Gordan coefficients.
To determine, for a given value of $B$, the allowed values of the $Y$, $I$ and $N$ quantum numbers
as well as which SU(3) irrep should be included in the basis we proceed as follows.
As already seen,  the value of the right hypercharge $Y_R \equiv - 2 F_8/\sqrt3$ is fixed by the constraint
$Y_R = B$. Thus, any SU(3) irrep that appears in the expansion,  Eq.(\ref{expansion}),  should have a maximum
value of hypercharge equal or larger than $B$. Thus, the possible values of $(p,q)$ should satisfy
\begin{equation}
{p + 2 q\over3} = B + m \ ,
\label{tril}
\end{equation}
with $p$ and $q$ non-negative integer numbers and $m=0,1,2,...$. The irreps corresponding to $m=0$
are the so-called minimal irreps that we will denote $(p_0, q_0)$. It is possible
to show\cite{Kop90} that the matrix element of $h$ in any state that belongs to a minimal
irrep is $<h>_0= 3 B/2$. To determine the relevant values of $Y$, $I$ and $N$ it is
enough to consider such irreps.  Although for non-vanishing strangeness  $S$ other values of the quantum
numbers could be allowed, they will be of no interest to us. In fact, it is not difficult to show
that for the first state with ``non-minimal quantum numbers" the matrix element of $h$ is more than
twice $<h>_0$. Therefore, such state is expected to appear as a highly excited
state in the spectrum.
Since the minimal irreps have maximum right hypercharge $Y_R = B$ it is clear
that corresponding possible values of the body-fixed isospin $N$ are $N = p_0/2$. On the other
hand, those of the hypercharge $Y$ are
\begin{equation}
- {2 p_0 + q_0\over3} \le Y \le {p_0 + 2 q_0\over3} \ .
\end{equation}
Finally, given a value of $Y$ that satisfies this relation, the allowed values of the isospin $I$
are
\begin{equation}
\left| {Y\over2} + {p_0-q_0\over3} \right| \le I
\le {p_0+q_0\over2} - {1\over2} \left| Y - {p_0-q_0\over3} \right| \ .
\end{equation}
In Table \ref{minirrep} we list, for each baryon number $3 \le B \le 9$, the minimal SU(3)
irrep which lead to states with $N < 3$ together with the allowed
values of isospin for some values of strangeness.
Given a set of possible $(B, I, Y, N)$ quantum numbers one should find
all the SU(3) irreps with $m > 0$ that enter in the expansion, Eq. (\ref{expansion}).
This is done by selecting from all the irreps which satisfy Eq. (\ref{tril})
those that contain a state with this same set of quantum numbers.
This leads to different towers of SU(3) irreps for each set of
quantum numbers. Once this is done, it is a simple task to transform
Eq. (\ref{eigen}) into an ordinary linear eigenvalue problem whose
solution provides the energy  eigenvalues $\epsilon$ and the coefficients
$\beta_{(p,q)}$. Of course, to do that one should work with a
basis of finite size. Since we are interested only in the few
lowest eigenvalues the minimum size is fixed by the
condition that those eigenvalues remain unchanged
under a further increase of such size.

Having determined $\Psi_{(Y, I, I_z),(B, N, N_3)}$ and the
corresponding possible quantum numbers we have still to obtain
the coefficients $\alpha^{JN}_{J_3 N_3}$ of
Eq. (\ref{jii}) and the allowed values of $J$.
For this purpose, only the spin $J$ and isospin $N$ are relevant.
Thus,  the situation is very similar to that of the $S=0$ case discussed
in Sec. IV of Ref.\cite{GSS00}. As already mentioned the full
wave function should transform as a one dimensional irrep of the
multisoliton symmetry group $G$. For the configurations we
are dealing with we have that, except for the
$B=5$ and $B=6$ cases, such one dimensional irrep is the
trivial irrep of the corresponding symmetry groups. For $B=5$,
$\Gamma$ is the $A_2$ irrep of $D_{2d}$,  while for
$B=6$ the wave functions should transform as the $A_2$ irrep
of $D_{4d}$. Using standard group theoretical arguments \cite{S93}
we know that the product representation $J \times N$
of SU(2) is in general a reducible representation of $G$.
The projector operator into the one dimensional irrep $\Gamma$ is
\begin{equation}
P_{\Gamma} = \frac{1}{|G|} \sum_{g \in G} \,
    \chi_{\Gamma}^*(g) \, \rho(g) \ ,
    \label{proj}
\end{equation}
where $|G|$ is the rank of the group, $\chi_{\Gamma}(g)$ the
character of operation $g$, and $\rho(g)$ the representation of $g$
in $J \times N$
\begin{equation}
\rho(g) = D^J(g) \times D^N(D_g) \ .
\end{equation}
where $D_g$ is the isospin operation associated with the space operation
$g$. The eigenvalues of $P_{\Gamma}$ can either vanish or be equal
to one. The eigenvectors corresponding to each non-vanishing
eigenvalue provide precisely the coefficients $\alpha^{JN}_{J_3 N_3}$
of Eq. (\ref{jii}), and there are as many wave functions as
non-zero eigenvalues. If all eigenvalues vanish there is no collective
state with the given $J,N$. If there is only one, the wavefunction is
an eigenfunction of the collective Hamiltonian. In case there would
be more than one, we choose those combinations which diagonalize
the parity operator.

\section{Numerical results}

To calculate the multibaryon spectra we  use the following set
of values for the parameters appearing in the effective action,
Eqs. (\ref{action}-\ref{oldmass}). We fix $f_\pi$, $m_\pi$ and
$m_K$ to their empirical values and take $e=4.1$ and
$f_K/f_\pi = 1.29$. This set of parameters leads to a single baryon
excitation spectrum which is in very good agreement with the one
observed for the octet and decuplet baryons. As well known, however,
the use of the empirical value for $f_\pi$ implies a $B=1$ Skyrmion mass
of around $1.7 \ {\rm GeV}$. Consequently, the absolute values of the calculated
masses come out to be too large. This problem is nowadays known to
be solved by the inclusion of Casimir effects\cite{MK91,Wal98}. We will
come back to this issue below.
With these values we can calculate $M_{sol}$ and the different quantities
that appear in the expression of $L_{coll}$ given by Eq. (\ref{lag}).
The results for the different baryon numbers up to $B=9$ are tabulated in
Tables \ref{tabmsol} and \ref{tabmom}. From Table \ref{tabmsol} we
observe that although $M_{sol}/B$ tends, on  average, to decrease as a
function of $B$ it always lies above $1.5 \ {\rm GeV}$. This clearly indicates that
Casimir effects will be also important to determine the absolute masses of
the configurations with $B > 1$. In any case, as in previous works where
$f_\pi$ was adjusted to reproduce the empirical nucleon mass, we observe
some deviation from a smooth behaviour. Also listed in Table
\ref{tabmsol} are the strange inertia parameter $K^S$ and the
symmetry breaking parameter $\gamma$. We see that, roughly, $K^S$ decreases
as $1/B$ while $\gamma$ increases as $B^2$. As we will see this
has important consequences on the amount of configuration
mixing as a function of the baryon number. In Table \ref{tabmom}
we list the spin, isospin and mixing inertia parameters for
the different values of $B$. We find that the values we obtained
behave, as a function of $B$, as those of Ref.\cite{GSS00}. In fact this
is to be expected since, as explained in that reference, such behaviour as well
as the number of independent components depends only on general properties of
the {\it Ans\"atze}.

Given the values of the inertia and symmetry breaking parameters
we can proceed to calculate the matrix elements of the rotational
Hamiltonian. For this purpose we have to find the solutions
of the eigenvalue equation Eq. (\ref{eigen}). As explained in the previous
section this amounts to determine the coefficients $\beta_{(p,q)}$
appearing in Eq. (\ref{expansion}). We have done this calculation for the
different sets of allowed quantum numbers. It interesting to note
that the amount of configuration mixing increases with $B$. This
can be clearly observed in Fig. 1 where we display the
decomposition of the lowest energy states with strangeness $S=0$
(full line) and $S=-B$ (dashed line) for $B=3$ and $B=9$.
In this figure the symbol $i$ labels  the different $(p,q)$ irrep
that appear in each decomposition. Of course, $i=0$ indicates
the corresponding  minimal irrep. We see that while for
$B=3$ about $80 \%$ of the wavefunction corresponds to the minimal
irrep, for $B=9$ such irrep represents less that $30 \%$ with the
rest of strength distributed in almost 10 irreps. This kind of
behaviour can be simply understood using second order perturbation theory. Within
that approximation $\beta_{i=1}$, that is the coefficient of the first
non-minimal irrep, will be proportional to $\gamma/( <h>_1 - <h>_0 )$.
It is not hard to show that, for the ground states with $S=0$, one has
$(<h>_1 - <h>_0)\propto B$. Since we have seen that $\gamma \propto B^2$
we obtain that $\beta_{i=1}$ should increase roughly as $B$.
Similar arguments can be used for the case $S=-B$. This explains
why the configuration mixing is quite independent of the value
of strangeness as it can be seen in Fig. 1 by comparing the solid
lines with the dashed ones. From the numerical point of view
the increase of configuration mixing implies that as larger
values of $B$ are considered one has to increase the size of the
basis in which the eigenfunction is expanded in order to
obtain convergence. In all the cases of interest we found that
no more than $15$ to $20$ configurations were needed.

The resulting multibaryon spectra are summarized in Tables IV and V.
In Table IV we report the rotational corrections to the masses of
the $S=0$ states. They are given as excitation energies taken with
respect to the corresponding lowest energy state whose absolute
rotational energy is indicated in brackets. It is important to
mention that for the $B=1$ systems it was shown
that this rather large absolute value is almost completely cancelled
by the Casimir corrections due to kaon loops\cite{Wal98}. Since similar
cancellations are expected to happen for $B >1$, the
excitation energies result to be the most meaningful quantities to look at.
We observe that the predicted spectra are in agreement with the ones obtained in the
alternative bound state approach \cite{GSS00}, except for a few
changes in the ordering of the states in the case of $S=-B$ and $B=5,8,9$.
 From the numbers presented in Tables II, IV and V it is apparent that
there is a clear separation of three different energy scales. There
is a 1 $\rm GeV$ scale related to the classical masses (per baryon number) and the eigenvalues
of Eq. (\ref{eigen}) for $S=0$ states,
there is another scale of about 300 $\rm MeV$ for the excitation of one unit  of strangeness,
and finally a 10-100 MeV scale related to
spin-isospin excitations. This last energy scale is evident in Table IV while it appears as
a small correction in Table V. In this way we recover the three leading order
contributions in the $N_c$ expansion $N_c , N_c^0$ and $N_c^{-1}$, which are more explicitly separated
in the BSA.

\section{Conclusions}

In this work we have studied the multibaryon spectra for baryon number $3\le B \le 9$
and strangeness values $S=0,-1,-B$ within the SU(3) collective coordinate
approach to the three flavor Skyrme model. To describe the classical background
solutions we have used {\it Ans\"atze} based on rational maps \cite{HMS98}, which provide very
good approximations  and also share the same symmetries as the exact solutions.
The symmetry structure is responsible for the spin and isospin assignments to
the spectrum states. Therefore, the collective Hamiltonians and wave functions
we obtain are of general validity, while the mass splittings depend on
the particular values of the moments of inertia and of the symmetry
breaking parameter.

We have found that, in general, the ordering of the different spin/isospin states
corresponding to a given baryon number as well as the energy separation between
those states obtained by using the present approach are very similar to the
results of the alternative bound state treatment of the SU(3) Skyrme model.
This fact together with the observation that in the collective approach the relative
strength of the flavor symmetry breaking term increases with increasing baryon number
(cf., Fig. 1) seems to indicate that both approaches tend to coincide as $B$ grows.
In this sense we can conclude that our finding that the increase of one unit
of strangeness implies a cost in energy of about 300 MeV rather independently of
$B \ge 3$ appears to be a rather general prediction of the SU(3) Skyrme model.

Finally, note that in the present calculation we have set the meson decay constants
to their empirical values. Consequently, all the resulting absolute masses are too large.
For example, we find values of $M_{sol}/B$ of about 1.60 GeV and  $S=0$ ground state
rotational corrections of about 0.8 GeV. These values are expected to be largely
compensated by the pion and kaon contributions to the Casimir energies, respectively.
In fact, this has been recently shown to happen in the $B=1$ sector of the model\cite{Wal98}. Unfortunately,
for $B > 1$ the difficulties associated with the treatment of the fluctuations around non-spherically
symmetric soliton backgrounds have prevented so far the explicit evaluation of the Casimir
effect even in the SU(2) case.

\acknowledgments

This work was supported in part by the grant PICT 03-00000-00133
from ANPCYT, Argentina. 
C.L.S. thanks FAPERJ for financial support and the kind
hospitality of Laboratorio TANDAR, CNEA, where part of this work was
done.

\appendix

\section*{}

In this Appendix we give the explicit expressions of the spin-isospin collective
Hamiltonian for $B \ge 3$. The form of these expressions depends only on
the symmetries of the soliton configurations. The method to derive them
is very similar to the one described in Sec. III of Ref. \cite{GSS00}. In fact,
the following expressions can be easily obtained from the ones given in that
reference by setting the corresponding hyperfine splitting constants to zero.

\begin{eqnarray}
H^{JN}_{B=3} &=& H^{JN}_{B=9} =
K^J \ {\hat J}^2 + K^N \ {\hat N}^2 - 2 K^M \ \vec {\hat N} \cdot {\vec{\hat J}} \ ,
\label{btres}
\\ \nonumber
\\
H^{JN}_{B=4} &=& K^J \ {\hat J}^2 + K_1^N \ {\hat N}^2 + (K_3^N - K_1^N) \ {\hat N_3}^2 \ ,
\label{bcuatro}
\\ \nonumber
\\
H^{JN}_{B=5} &=& K_1^J \ ( {\hat J}^2 - {\hat J_3}^2 ) + K_1^N \ ( {\hat N}^2 - {\hat N_3}^2 ) -
2 K_1^M \ ( \vec {\hat N} \cdot \vec {\hat J} - \hat N_3 \ \hat J_3 )
\nonumber \\
& & + K_3^J \ \hat J_3^2 + K_3^N \ \hat N^2_3 - 2 K_3^M  \ \hat N_3 \ \hat J_3  \ ,
\label{bcinco}
\\ \nonumber
\\
H^{JN}_{B=6} &=& H^{JN}_{B=8} = K_1^J \ \hat J^2 +  K_1^N \ \hat N^2
+ (K_3^J - K_1^J) \ \hat J_3^2 + (K_3^N - K_1^N) \ \hat N^2_3 - 2 K_3^M \ \hat N_3 \ \hat J_3
\label{bseis} \ ,
\\ \nonumber
\\
H^{JN}_{B=7} &=& K^J \ \hat J^2 + K^N \ \hat N^2 \ .
\end{eqnarray}

\begin{table}
\narrowtext

\caption{Minimal SU(3) irreps and allowed values of $N$ and $I$ for states with some selected
values of strangeness for $B=3-9$. Only states with $N < 3$ are listed.}
\label{minirrep}

\begin{center}
\begin{tabular}{cccccc}
$B$    & Minimal SU(3) irrep  & $N$  & \multicolumn{3}{c}{Allowed values of $I$} \\ \cline{4-6}
       &                        &      &  $ S=0 $   & $ S=-1 $      & $ S = -B $   \\
\hline
3      & ${\bf{\overline{ 35}}}$       & 1/2  &   1/2    & 0, 1       & 1,2      \\
       & ${\bf            64  }$       & 3/2  &   3/2    & 1, 2       & 0,1,2,3      \\
       & ${\bf            81  }$       & 5/2  &   5/2    & 2, 3       & 1,2,3      \\
4      & ${\bf{\overline{ 28}}}$       & 0    &   0      & 1/2        & 2         \\
       & ${\bf{\overline{ 81}}}$       & 1    &   1      & 1/2, 3/2   & 1,2,3     \\
       & ${\bf           125  }$       & 2    &   2      & 3/2, 5/2   & 0,1,2,3,4 \\
5      & ${\bf{\overline{ 80}}}$       & 1/2  &  1/2     & 0, 1       & 2,3        \\
       & ${\bf{\overline{154}}}$       & 3/2  &  3/2     & 1, 2       & 1,2,3,4     \\
       & ${\bf            216 }$       & 5/2  &  5/2     & 2, 3       & 0,1,2,3,4,5 \\
6      & ${\bf{\overline{ 55}}}$       & 0    &  0       & 1/2        & 3           \\
       & ${\bf{\overline{162}}}$       & 1    &  1       & 1/2, 3/2   & 2,3,4       \\
       & ${\bf{\overline{260}}}$       & 2    &  2       & 3/2, 5/2   & 1,2,3,4,5   \\
7      & ${\bf{\overline{143}}}$       & 1/2  &  1/2     & 0, 1       & 3,4           \\
       & ${\bf{\overline{280}}}$       & 3/2  &  3/2     & 1, 2       & 2,3,4,5       \\
       & ${\bf{\overline{405}}}$       & 5/2  &  5/2     & 2, 3       & 1,2,3,4,5,6 \\
8      & ${\bf{\overline{ 91}}}$       & 0    &  0       & 1/2        & 4           \\
       & ${\bf{\overline{270}}}$       & 1    &  1       & 1/2, 3/2   & 3,4,5       \\
       & ${\bf{\overline{440}}}$       & 2    &  2       & 3/2, 5/2   & 2,3,4,5,6   \\
9      & ${\bf{\overline{224}}}$       & 1/2  &  1/2     & 0, 1       & 4,5           \\
       & ${\bf{\overline{442}}}$       & 3/2  &  3/2     & 1, 2       & 3,4,5,6       \\
       & ${\bf{\overline{648}}}$       & 5/2  &  5/2     & 2, 3       & 2,3,4,5,6,7
\end{tabular}
\end{center}

\end{table}

\begin{table}
\mediumtext

\caption{Soliton mass (per baryon unit), strangeness inertia parameter and symmetry breaking
parameter for $B=3-9$.}
\label{tabmsol}
\begin{center}
\begin{tabular}{cccc}
 B  & $M_{sol}/B$   &  $K^S$   &  $\gamma$  \\
    &    $({\rm GeV})$      &  $(\rm{MeV})$   &            \\ \hline
 3  &  1.64       &  55.12   &  38.43    \\
 4  &  1.58       &  43.18   &  61.18    \\
 5  &  1.59       &  32.80   & 105.69    \\
 6  &  1.58       &  27.03   & 154.83    \\
 7  &  1.54       &  23.96   & 194.76    \\
 8  &  1.56       &  20.04   & 279.99    \\
 9  &  1.57       &  17.25   & 379.42    \\
\end{tabular}
\end{center}

\end{table}

\begin{table}
\mediumtext

\caption{Spin, isospin and mixed inertia parameters for $B=3-9$.}

\label{tabmom}

\begin{center}
\begin{tabular}{cccc}
$B$  & $K^J ~(\rm{MeV})$ & $K^N ~(\rm{MeV})$ &   $K^M ~(\rm{MeV})$ \\ \hline
3         &   11.28    &   37.99   &    7.19  \\
4         &    6.29    &   28.94    &     0     \\
          &            &   28.94    &           \\
          &            &   24.10    &           \\
5         &   3.77     &   20.74    &   -0.88   \\
          &   3.77     &   20.74    &   -0.88   \\
          &   4.27     &   24.71    &   -0.67   \\
6         &   2.66     &   19.06    &    0      \\
          &   2.66     &   19.06    &    0      \\
          &   3.09     &   17.93    &   0.94    \\
7         &   2.23     &   16.72    &    0      \\
8         &   1.73     &   14.23    &    0      \\
          &   1.73     &   14.23    &    0      \\
          &   1.53     &   15.38    &  -0.47    \\
9         &   1.29     &   13.03    &  -0.33    \\
\end{tabular}
\end{center}

\end{table}

\begin{table}
\caption{Quantum numbers and rotational excitation energies for the $S=0$ states.
The excitation energies are taken with respect to that of the lowest energy state
for each baryon number. The absolute rotational energies of those
states are indicated in brackets.}

\label{nonstr}

\begin{center}
\begin{tabular}{ccccc}
$B$    & $J^P$    &   $I$   &  $N$    &$E_{exc} (\rm{MeV})$ \\  \hline
 3     &${1/2}^+$ & ${1/2}$ &   1/2   & (847)  \\
       &${5/2}^-$ & ${1/2}$ &   1/2   &  61    \\
       &${3/2}^-$ & ${3/2}$ &   3/2   & 110    \\
 4     &$ 0^+   $ &    0    &   0     & (808)  \\
       &$ 4^+   $ &    0    &   0     & 126    \\
       &$ 0^+   $ &    2    &   2     & 180    \\
 5     &${1/2}^+$ & ${1/2}$ &   1/2   & (837)  \\
       &${3/2}^+$ & ${1/2}$ &   1/2   &  9     \\
       &${3/2}^-$ & ${1/2}$ &   1/2   &  11    \\
 6     &$ 1^+   $ &    0    &   0     & (827)  \\
       &$ 3^+   $ &    0    &   0     &  27    \\
       &$ 0^+   $ &    1    &   1     &  34    \\
 7     &${7/2}^+$ &${1/2}$  &   1/2   & (872)  \\
       &${3/2}^+$ &${3/2}$  &   3/2   &  24    \\
       &${9/2}^+$ &${3/2}$  &   3/2   &  71    \\
 8     &$  0^+  $ &   0     &   0     & (828)  \\
       &$  2^+  $ &   0     &   0     &  10    \\
       &$  1^+  $ &   1     &   1     &  32    \\
 9     &$  1/2^+$ &   1/2   &   1/2   & (842)  \\
       &$  5/2^-$ &   1/2   &   1/2   &   12   \\
       &$  7/2^-$ &   1/2   &   1/2   &   18   \\
\end{tabular}
\end{center}


\end{table}

\begin{table}


\mediumtext \caption{Quantum numbers and rotational excitation energies
(per unit of strangeness) for $S=-1$ and $S=-B$ states.
The excitation energies (in {\rm MeV}) are taken with respect to that of the $S=0$ lowest
energy state for each baryon number.
The absolute rotational energies of those states are given in Table IV.}

\label{str}

\begin{center}
\begin{tabular}{ccccccccc}
  &\multicolumn{4}{c}{$S=-1$} & \multicolumn{4}{c} {$S=-B$}  \\
\cline{2-5}
\cline{6-9}
$B$    & $J^P$    &   $I$   &  $N$  &$E_{exc}/|S|$&
         $J^P$    &   $I$   &  $N$  &$E_{exc}/|S|$ \\  \hline
 3     &${1/2}^+$ &    0    &${1/2}$& 263.5  &
        ${1/2}^+$ &    1    &${1/2}$& 291.7 \\
       &${1/2}^+$ &    1    &${1/2}$& 304.0  &
        ${3/2}^-$ &    0    &${3/2}$& 292.9 \\
       &${5/2}^-$ &    0    &${1/2}$& 325.0  &
        ${5/2}^+$ &    0    &${3/2}$& 304.4 \\
 4     &$ 0^+   $ & ${1/2}$ &   0   & 287.9  &
        $ 0^+   $ &    0    &   2   & 302.8 \\
       &$ 4^+   $ & ${1/2}$ &   0   & 413.7  &
        $ 0^+   $ &    2    &   0   & 308.5 \\
       &$ 0^+   $ & ${3/2}$ &   2   & 425.1  &
        $ 0^+   $ &    1    &   2   & 311.8 \\
 5     &${1/2}^+$ &    0    &${1/2}$& 279.4  &
        ${1/2}^+$ &    1    &${3/2}$& 301.7 \\
       &${3/2}^+$ &    0    &${1/2}$& 288.1  &
        ${1/2}^-$ &    1    &${3/2}$& 303.6 \\
       &${3/2}^-$ &    0    &${1/2}$& 290.8  &
        ${3/2}^-$ &    1    &${3/2}$& 304.7 \\
 6     &$ 1^+   $ & ${1/2}$ &$  0  $& 299.1  &
        $ 0^-   $ &    1    &   2  & 308.6 \\
       &$ 0^+   $ & ${1/2}$ &$  1  $& 313.5  &
        $ 1^-   $ &    1    &$  2  $& 309.5 \\
       &$ 3^+   $ & ${1/2}$ &$  0  $& 325.7  &
        $ 1^+   $ &    1    &$  2  $& 310.3 \\
 7     &${7/2}^+$ &   0   &${1/2}$&  282.0  &
        ${3/2}^+$ &   2   &${3/2}$&  301.3 \\
       &${3/2}^+$ &   1   &${3/2}$&  298.7  &
        ${5/2}^+$ &   1   &${5/2}$&  302.8 \\
       &${7/2}^+$ &   1   &${1/2}$&  299.1  &
        ${7/2}^+$ &   1   &${5/2}$&  305.1 \\
 8     &$  0^+  $ &${1/2}$&   0   &  301.3  &
        $  0^+  $ &   2   &   2   &  313.5 \\
       &$  2^+  $ &${1/2}$&   0   &  311.7  &
        $  1^+  $ &   1   &   3   &  314.8 \\
       &$  1^+  $ &${1/2}$&   1   &  319.3  &
        $  2^-  $ &   2   &   2   &  314.8 \\
 9     &$  1/2^+    $ &   0     &    1/2  &  296.5  &
        $  1/2^-    $ &   2     &    5/2  &  318.0 \\
       &$  5/2^-    $ &   0     &    1/2  &  308.1  &
        $  3/2^-    $ &   2     &    5/2  &  318.3 \\
       &$  1/2^+    $ &   1     &    1/2  &  309.4  &
        $  5/2^+    $ &   2     &    5/2  &  318.4 \\
\end{tabular}
\end{center}
\end{table}

\begin{figure}[p]
\centerline{\psfig{figure=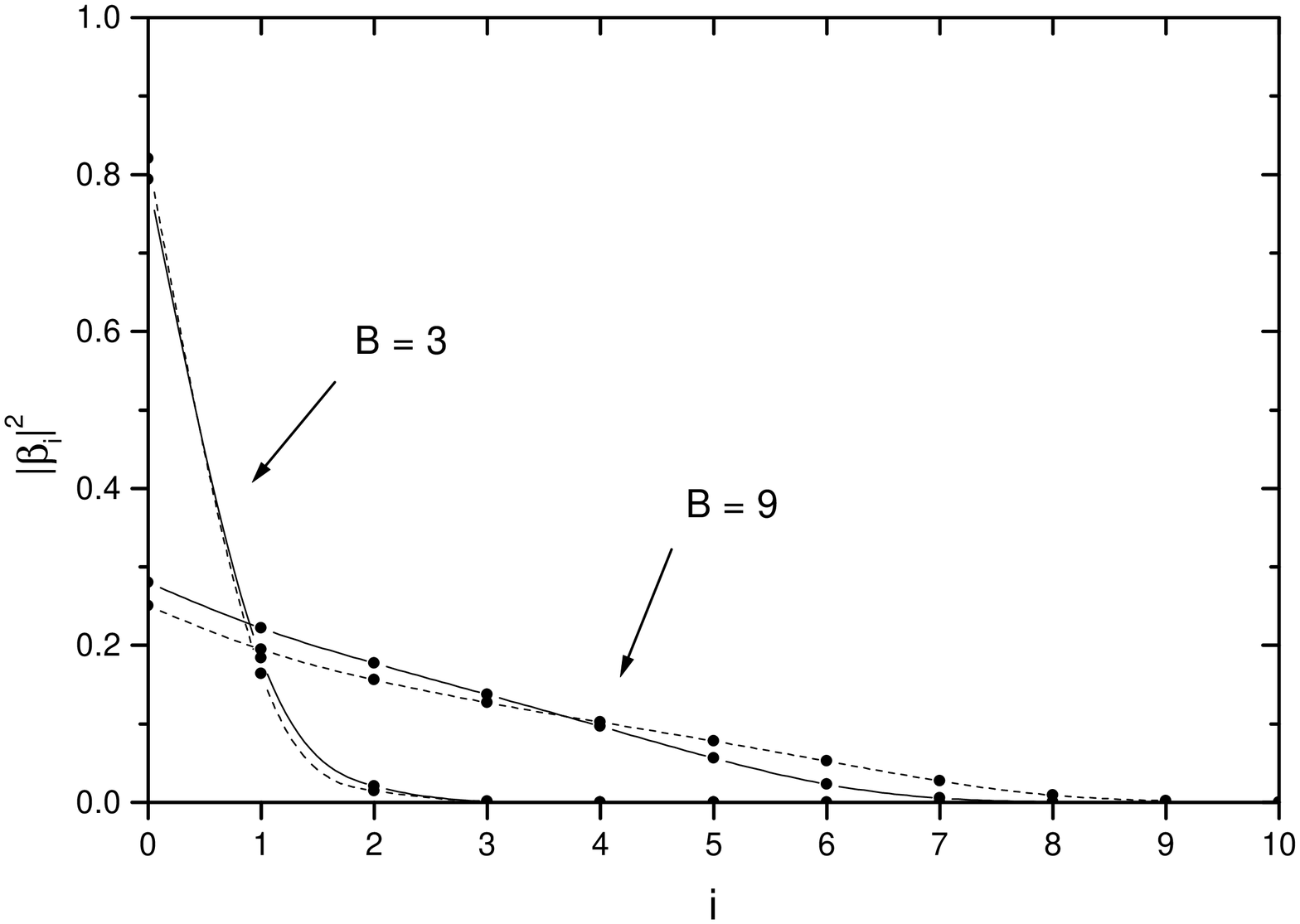,height=13.cm}} \caption[]
{Contribution of higher irreps to the lowest energy states
with $S=0$ (full line) and $S=-B$ (dashed line). }
\label{f1}
\end{figure}

\begin{thebibliography}{99}

\bibitem{Sky61}
T. H.\ R.\ Skyrme,
Proc.\ Roy.\ Soc.\ (London) {\bf 260}, 127 (1961);
Nucl.\ Phys.\ {\bf 31}, 556 (1962).

\bibitem{ZB86}
For early reviews on the SU(2) Skyrme model see:\\
G.~Holzwarth and B.~Schwesinger,
Rep. Prog. Phys. {\bf 49}, 825 (1986);\\
I.~Zahed and G.~E. Brown,
Phys. Rep. {\bf 142}, 481 (1986).

\bibitem{Wei96}
For a rather recent review on the extensions of
the model to flavor SU(3) see:\\
H. Weigel,
Int J. Mod. Phys. A {\bf 11}, 2419 (1996).

\bibitem{KS87}
V. B. Kopeliovich and B. E. Stern,
JETP Lett. {\bf 45}, 203 (1987);
J. J. M. Verbaarschot,
Phys. Lett. B {\bf 195}, 235 (1987);
N. S. Manton,
{\it ibid.} {\bf 192}, 177 (1987).

\bibitem{BTC90}
E. Braaten, S. Townsend, and L. Carson,
Phys. Lett. B {\bf 235}, 147 (1990).

\bibitem{BS97}
R. A. Battye and P. M. Sutcliffe,
Phys. Rev. Lett. {\bf 79}, 363 (1997).

\bibitem{HMS98}
C. J. Houghton, N. S. Manton,  and P. M. Sutcliffe,
Nucl. Phys. {\bf B510}, 507 (1998).

\bibitem{BBT97}
C. Barnes, W. K. Baskerville,  and N. Turok,
Phys. Rev. Lett. {\bf 79}, 367 (1997);
Phys. Lett. B {\bf 411}, 180 (1997);
W. K. Baskerville, `` Vibrational spectrum of the B=7 Skyrme soliton,''
 hep-th/9906063.

\bibitem{SS98}
M. Schvellinger and N. N. Scoccola,
Phys. Lett. B {\bf 430}, 32 (1998).

\bibitem{GSS00}
J. P. Garrahan, M. Schvellinger,  and N. N. Scoccola,
Phys. Rev. D {\bf 61}, 014001 (2000).

\bibitem{SS00}
C. L. Schat and N. N. Scoccola,
Phys. Rev. D {\bf 61}, 034008 (2000).

\bibitem{E864}
K. Borer {\it et al.} ,
Phys. Rev. Lett. {\bf 72}, 1415 (1994);
T. A. Armstrong et al. (E864 experiment),
Phys. Rev. Lett. {\bf 79}, 3612 (1997).

\bibitem{SV98}
J. Schaffner-Bielich and A. P. Vischer,
Phys. Rev. D {\bf 57}, 4142 (1998).

\bibitem{CK85}
C. G. Callan and I. Klebanov,
Nucl. Phys. {\bf B262}, 365 (1985);
N. N. Scoccola, H. Nadeau, M. Nowak,  and M. Rho,
Phys. Lett. B {\bf 201}, 425 (1988);
C. G. Callan, K. Hornbostel, and I. Klebanov,
{\it ibid.} {\bf 202}, 269 (1988);
J. P. Blaizot, M. Rho, and N. N. Scoccola,
{\it ibid.} {\bf 209}, 27 (1988);
U. Blom, K. Dannbom, and D. O. Riska,
Nucl. Phys. {\bf A493}, 384 (1989).

\bibitem{YA88}
H. Yabu and K. Ando,
Nucl. Phys. {\bf B301}, 601 (1988).


\bibitem{Kop90}
V. Kopeliovich,
Phys. Lett. B {\bf 259}, 234 (1990).

\bibitem{S93}
J.-P. Serre, {\it Linear Representations of Finite Groups},
(Springer, New York, 1993).

\bibitem{MK91}
B. Moussallam and D. Kalafatis,
Phys. Lett. B {\bf 272}, 196 (1991);
B. Moussallam,
Ann. Phys. (N.Y.) {\bf 225}, 264 (1993);
F. Meier and H. Walliser,
Phys. Rep. {\bf 289}, 383 (1997).


\bibitem{Wal98}
H. Walliser,
Phys. Lett. B {\bf 432}, 15 (1998);
N. N. Scoccola and H. Walliser,
Phys. Rev. D {\bf 58}, 094037 (1998).


\end{thebibliography}
\end{document}